\documentclass[12pt]{article}

\usepackage{graphicx}
\usepackage{amsmath}
\usepackage{amsfonts}
\usepackage{amssymb}
\usepackage{bbm}
\usepackage{hyperref}

\begin{document}

\newcommand{\ba}[1]{\begin{array}{#1}} \newcommand{\ea}{\end{array}}
\newcommand{\cleqn}{\setcounter{equation}{0}}

\numberwithin{equation}{section}


\def\Journal#1#2#3#4{{#1} {\bf #2}, #3 (#4)}

\def\NCA{\em Nuovo Cimento}
\def\NIM{\em Nucl. Instrum. Methods}
\def\NIMA{{\em Nucl. Instrum. Methods} A}
\def\NPB{{\em Nucl. Phys.} B}
\def\PLB{{\em Phys. Lett.}  B}
\def\PRL{\em Phys. Rev. Lett.}
\def\PRD{{\em Phys. Rev.} D}
\def\ZPC{{\em Z. Phys.} C}

\def\st{\scriptstyle}
\def\sst{\scriptscriptstyle}
\def\mco{\multicolumn}
\def\epp{\epsilon^{\prime}}
\def\vep{\varepsilon}
\def\ra{\rightarrow}
\def\ppg{\pi^+\pi^-\gamma}
\def\vp{{\bf p}}
\def\ko{K^0}
\def\kb{\bar{K^0}}
\def\al{\alpha}
\def\ab{\bar{\alpha}}

\def\np{Nucl. Phys. {\bf B}}\def\pl{Phys. Lett. {\bf B}}
\def\mpl{Mod. Phys. {\bf A}}\def\ijmp{Int. J. Mod. Phys. {\bf A}}
\def\cmp{Comm. Math. Phys.}\def\prd{Phys. Rev. {\bf D}}

\def\oa{\bigcirc\!\!\!\! a}
\def\ob{\bigcirc\!\!\!\! b}
\def\oc{\bigcirc\!\!\!\! c}
\def\oi{\bigcirc\!\!\!\! i}
\def\oj{\bigcirc\!\!\!\! j}
\def\ok{\bigcirc\!\!\!\! k}
\def\ve{\vec e}\def\vk{\vec k}\def\vn{\vec n}\def\vp{\vec p}
\def\vr{\vec r}\def\vs{\vec s}\def\vt{\vec t}\def\vu{\vec u}
\def\vv{\vec v}\def\vx{\vec x}\def\vy{\vec y}\def\vz{\vec z}

\def\ve{\vec e}\def\vk{\vec k}\def\vn{\vec n}\def\vp{\vec p}
\def\vr{\vec r}\def\vs{\vec s}\def\vt{\vec t}\def\vu{\vec u}
\def\vv{\vec v}\def\vx{\vec x}\def\vy{\vec y}\def\vz{\vec z}

\newcommand{\AdS}{\mathrm{AdS}}
\newcommand{\dd}{\mathrm{d}}
\newcommand{\eee}{\mathrm{e}}
\newcommand{\sgn}{\mathop{\mathrm{sgn}}}

\def\a{\alpha}
\def\b{\beta}
\def\g{\gamma}

\newcommand\lsim{\mathrel{\rlap{\lower4pt\hbox{\hskip1pt$\sim$}}
    \raise1pt\hbox{$<$}}}
\newcommand\gsim{\mathrel{\rlap{\lower4pt\hbox{\hskip1pt$\sim$}}
    \raise1pt\hbox{$>$}}}

\newcommand{\beq}{\begin{equation}}
\newcommand{\eeq}{\end{equation}}
\newcommand{\bea}{\begin{eqnarray}}
\newcommand{\eea}{\end{eqnarray}}
\newcommand{\bem}{\begin{pmatrix}}
\newcommand{\eem}{\end{pmatrix}}
\newcommand{\noi}{\noindent}


\begin{flushright}
October, 2013
\end{flushright}

\bigskip

\begin{center}

{\Large\bf  Scalar potentials, propagators and global symmetries in AdS/CFT}


\vspace{1cm}

\centerline{
Borut Bajc$^{a,b,}$\footnote{borut.bajc@ijs.si} and
Adri\'{a}n R. Lugo$^{a,c,}$\footnote{lugo@fisica.unlp.edu.ar}
}
\vspace{0.5cm}
\centerline{$^{a}$ {\it\small J.\ Stefan Institute, 1000 Ljubljana, Slovenia}}
\centerline{$^{b}$ {\it\small Department of Physics, University of Ljubljana, 1000 Ljubljana,
Slovenia}}
\centerline{$^{c}$ {\it\small Instituto de F\'\i sica de La Plata-CONICET,
and Departamento de F\'\i sica,}}
\centerline{ {\it\small Facultad de Ciencias Exactas, Universidad Nacional de La Plata,
Argentina}\footnote{Permanent address.}}

\end{center}

\bigskip

\begin{abstract}

We study the transition of a scalar field in a fixed $AdS_{d+1}$ background between an extremum and a
minimum of a  potential. We first prove that two conditions must be met for the solution to exist. First, the
potential involved cannot be generic, i.e. a fine-tuning of their parameters is mandatory. Second, at least in
some region its second derivative must have a negative upper limit which depends only on the dimensionality $d$.
We then calculate the boundary propagator for small momenta in two different ways: first in a WKB approximation,
and second with the usual matching method, generalizing the known calculation to arbitrary order. Finally, we
study a system with spontaneously broken non-Abelian global symmetry, and show in the holographic language
why the Goldstone modes appear.

\end{abstract}

\clearpage

\tableofcontents


\section{Introduction}

The simplest example of AdS-CFT correspondence \cite{Maldacena:1997re,Gubser:1998bc,Witten:1998qj}
is gravity plus a real scalar field system in asymptotic AdS (for a partial list see
\cite{Girardello:1998pd,Freedman:1999gk,Arutyunov:2000rq,Muck:2001cy,Martelli:2001tu,Berg:2002hy,Freedman:2003ax}).

Apart from some special cases (see for example \cite{Cvetic:1992bf}) it is expected that the even simplified version of such systems, i.e. the no-back-reaction limit where the gravitational coupling $\kappa\rightarrow0$,
would give the relevant information (for some reviews on this subject see
for example \cite{Aharony:1999ti,DeWolfe:2000xi,D'Hoker:2002aw,Skenderis:2002wp}).
Recently this has been done in \cite{Bajc:2012vk}, where the potential of the real scalar field has been approximated by a piece-wise quadratic potential in order to allow analytic treatment. It has been then shown that: a) in order for the solution between the UV extremum and the IR minimum to exist, there must be some non-trivial constraint among parameters in the potential; b) at least one region needs $V''<-d^2/4$; c) a solution of such a system has vanishing action and d) the propagator in the boundary theory exhibits a simple $1/q^2$ pole as predicted by the Goldstone theorem applied to the spontaneously broken dilatation invariance \cite{Bianchi:2001de}.

The last two points has been considered in more detail in \cite{Bajc:2013wha} (see also \cite{Hoyos:2013gma} and
\cite{Kol:2013msa}) following
an inspiring paper \cite{Hoyos:2012xc}, where it was explicitly shown that even such a simplified system has a BPS
type solution which exhibits the Goldstone theorem for a spontaneously broken conformal invariance in subtle way, i.e.
mixing the normalizable and non-normalizable modes in the bulk at the next-to-leading order of the matching method.

The purpose of this contribution is twofold. First, we would like to shed more light on the first two issues, i.e. on the constraints
the potential must satisfy for allowing a solution. Second, we would like to see the $1/q^2$ propagator of the Goldstone
in a different way, generalize the matching method at all orders, and present few examples of its use.

The plan of the paper is the following. After setting the notation and main formulae in Section \ref{kappa0},we summarize in
Section \ref{analytic} the wall solution found in \cite{Bajc:2012vk}. In Section \ref{generic} we then explain the reason
for a fine-tuning of the potential parameters and explicitly show how one can find BPS-type solutions to the first order
equation of motion even in this no-backreaction limit, i.e. generalize the usual $\kappa\ne0$ expression of the potential
through the superpotential to the $\kappa\to0$ limit. With it we can prove in Section \ref{mass} in complete generality that
for the solution to exist, the second derivative of the scalar potential must be smaller than $-d^2/4$ in at least some region.
In section \ref{WKB} we find the same $1/q^2$ propagator in the $q\to0$ limit of the dilaton using then WKB approximation,
while a long Section \ref{allorders} is devoted to a detailed analysis of the matching method to all orders. This is then used in
Section \ref{global} to show in an explicit example what exactly makes the Goldstone boson of a global symmetry massless
in the holographic language.

\section{\label{kappa0}The no back-reaction limit ($\kappa\to0$)}

We will consider in most of this paper a real scalar field $t$ in $d+1$ dimensions with bulk euclidean action
\beq
\label{actionphi}
S^{(bulk)}[t] =  \int d^{d+1}x\,\sqrt{\det{g_{ab}}}\;
\left(\frac{1}{2}\,g^{ab}\;\partial_at\; \partial_bt+V(t)\right)
\eeq
in a non-dynamical $AdS_{d+1}$ background
\beq\label{metric}
g = \frac{1}{z^2}\left(dz^2+ \delta_{\mu\nu}\;dx^\mu\;dx^\nu\right)
\eeq
where $(x^\mu)$ are the QFT coordinates with $x^d\equiv i\,x^0$ the euclidean time and the
AdS scale has been set to $1$. The boundary is located at $z=0$ (UV region) while the horizon
is at $z=\infty$ (IR region).

The dimensionless field variable $t$ is normalized to have extrema of the potential at $t=0,1$.
More precisely, we will consider potentials ( throughout the paper
we will indicate with a dot the derivative w.r.t. the bulk coordinate $z$ and with a prime a field derivative)
\beq\label{condpot}
V(0)=0\quad,\quad V'(0)=0\qquad;\qquad V(1)<0\quad,\quad V'(1)=0\quad,\quad V''(1)>0
\eeq
i.e. $t=1$ will be the true minimum, while at the origin the potential can
have a minimum (being a false vacuum thus) or even a maximum, provided that it is in the
Breitenl\"ohner-Freedman conformal window $-d^2/4<V''(0)<0$.

We will be interested in regular, Poincar\`e invariant solutions $t=t(z)$
that interpolate between the UV and IR regions.
They obey the equation of motion
\beq\label{eom}
z^2\;\ddot{t}(z) -(d-1)\;z\; \dot{t}(z) = V'(t)
\eeq
and necessary behave in the UV and IR as
\beq\label{bc}
t(z)\xrightarrow{z\rightarrow 0} a_{UV}\; z^{\Delta^{UV}}
\qquad;\qquad t(z)\xrightarrow{z\rightarrow\infty} 1 + a_{IR}\; z^{d-\Delta^{IR}}
\eeq
respectively, where
\beq\label{Delta}
\Delta^{UV/IR} \equiv \frac{d}{2} + \nu_{UV/IR}\qquad;\qquad
\nu_{UV/IR}\equiv\sqrt{\frac{d^2}{4} + m^2_{UV/IR}}
\eeq
with $m^2_{UV}\equiv V''(0)$ and $m^2_{IR}\equiv V''(1)>0$ ($t=1$ is a minimum according to
(\ref{condpot}))\footnote{\label{foot1}In the window $-\frac{d^2}{4}<V''(0)<0$ the term
$z^{d-\Delta^{UV}}$ could also be present in the small $z$ power expansion of $t(z)$.
From the AdS/CFT point of view this term is interpreted as a source that breaks
explicitly the scale invariance of the boundary QFT; then we should not expect a
Goldstone mode to appear, situation we are not interested in.
These domain walls are interpreted as dual to renormalization group flows generated by
deformation of the UV CFT by a relevant operator, i.e. one of dimension less than $d$ \cite{Skenderis:2002wp}.
}.

We recall as a last remark that the symmetries of $AdS$ space translate in the scale invariance of equation
(\ref{eom}), i.e. if $t(z)$ is a solution so it is $t(\lambda z)$, a fact of great relevance.

\section{\label{analytic}Analytic solutions for approximated bulk potentials}

In this section we shortly summarize the results presented in \cite{Bajc:2012vk}.

The interesting region for $t$ is between the local minimum at $0$ and
the global minimum at $1$.
We will divide this region into a number of sections, and in each of them the potential can
be locally approximated by a quadratic form:
\beq
\label{vt}
V(t)=\frac{A}{2}\;t^2+B\;t+C
\eeq
The minimum number of such sections is three: (1) $0<t<t_1$, (2) $t_1<t<t_2$, (3) $t_2<t<1$.
The coefficients in (\ref{vt}) are parameterized in each region as
\bea
\label{ABC}
A &=& \left\{\begin{array}{r}A_1>0\\A_2<0\\A_3>0 \end{array}\right.
\qquad;\qquad B = \left\{ \begin{array}{l} 0\\-A_2\,t_M\\-A_3\end{array}\right.
\qquad;\qquad\cr
C &=& \left\{ \begin{array}{l}0\\(A_1-A_2)\;t_1{}^2/2+A_2\;t_M\;t_1\cr
(A_2-A_3)\;t_2{}^2/2+A_3\;t_2-A_2\;t_M\;t_2+(A_1-A_2)\;t_1{}^2/2+A_2\;t_M\;t_1 \end{array}\right.
\eea
respectively.
The strange choice of $C$'s is required by the continuity of the potential.
Furthermore, we will require the continuity of the first derivatives of the
potential which yields to
\beq\label{contderpot}
t_1 = \frac{-A_2\;t_M}{A_1 - A_2}\qquad;\qquad
t_2 = \frac{A_3-A_2\,t_M}{A_3-A_2}
\eeq
relations that automatically satisfy $0<t_1<t_M<t_2<1$ for any $0<t_M<1$.
In this way we remain with four relevant parameters, the $A_i$'s and $t_M$.

Similarly as in (\ref{Delta}) we introduce
\beq\label{mus}
\Delta^\pm_i=d/2\pm\nu_i\qquad;\qquad
\nu_i{}^2 \equiv \frac{d^2}{4}+ A_i
\eeq
We will consider the case of real $\nu_{1,3}>\frac{1}{2}$ ($A_{1,3}>0$) and pure imaginary
$\nu_2\equiv i\,\bar\nu_2$ ($A_2<-\frac{d^2}{4}$) with $\bar\nu_2>0$.

The solution to (\ref{eom}) with the piece-wise quadratic potential (\ref{vt}) is known

\beq
\label{tb}
t_b(z)=\left\{
\begin{array}{lcr}
t_1\; (z/z_1)^{\Delta_1^+} & , &  0<z<z_1\\
t_M+D_+\,(z/z_2)^{\Delta_2^+}+D_-\,(z/z_2)^{\Delta_2^-}& , & z_1<z<z_2\\
1-(1-t_2)\,(z/z_2)^{\Delta_3^-} & , &  z_2<z<\infty
\end{array}
\right.
\eeq

Continuity of the solution and its derivative at $z_{1,2}$ requires

\bea
D_+&=&\frac{(1-t_M)\Delta_2^-\Delta_3^-(\nu_3-i\bar\nu_2)}{2i\bar\nu_2(A_3-A_2)}
\hskip 0.5cm,\hskip0.5cm D_-=D_+^*\\
\label{tm}
t_M&=&\left(1-\frac{\Delta_1^+}{\Delta_3^-}\left(\frac{\nu_3^2+\bar\nu_2^2}{\nu_1^2+\bar\nu_2^2}\right)^{1/2}
\left(\frac{z_2}{z_1}\right)^{d/2}\right)^{-1}\\
\label{z2z1}
\bar\nu_2\log{(z_2/z_1)}&=&(2k+1)\pi-\alpha_1-\alpha_3
\eea

\noi
with

\beq
\alpha_i\equiv\arctan{(\bar\nu_2/\nu_i)}\hskip0.5cm,\hskip0.5cm i=1,3
\eeq

Notice here two things:

\begin{itemize}

\item
there is one relation (fine-tuning) among the potential parameters $A_i$, $t_M$, see
eqs. (\ref{tm}) and (\ref{z2z1}),

\item
$\nu_2$ needs to be purely imaginary.

\end{itemize}

The whole procedure can be repeated with more intervals, but these two conclusions still remain:
A non-trivial fine-tuning among parameters is needed, and at least in one interval $\nu^2=d^2/4+V''$
needs to be negative. In the next two sections we will try to understand better these two issues.

\section{\label{generic}Why the potential cannot be generic}

Eqs. (\ref{tm}) and (\ref{z2z1}) represent the quantization condition on the potential for the
solution to exist at all.
As it has been noted in \cite{Bajc:2012vk} and remarked before, this follows from the invariance of
the equation of motion under dilatations $z\to\lambda z$ for any positive real $\lambda$.
There is thus an infinite family of solutions: the location of the domain wall is
not determined. In our previous example this is seen explicitly by the fact that the coordinates
$z_1$ and $z_2$ cannot be determined both, but due to dilatation invariance of the original
equation of motion only their ratio. The four equations (functions and derivatives at $z_{1,2}$)
cannot be satisfied by only three parameters $D_{+,-}$, $z_2/z_1$, so a non-trivial relation
among potential parameters follow. This simple counting can be easily generalized to an arbitrary
number of intervals.

What happens if a fine-tuned potential changes a bit, i.e. if we relax the constraint among the potential parameters?
The numerical output will make $t(z)$ diverge, so that for $z\to\infty$ limit it will not reach the unit value. In other words, the transition is not from the extremum in the origin to the minimum at $t=1$, but it escapes to infinity.
In order to make the field land to the minimum, one needs a constrained value for the model parameters.

There are two simple ways to see why there must be some constraint among the model parameters,
if we are looking for a solution of (\ref{eom}).

First of all, we have a second order differential equation. In the limit $z\to0$ this non-linear equation can be linearized,
call the two independent solutions of this linearized version $t_+(z)$ and $t_-(z)$. Let they be defined so that for $z\to0$,
$t_+(z)\propto z^{\Delta^{UV}}$ with $\Delta^{UV}$ given in (\ref{Delta}) and $t_-(z)\propto z^{d-\Delta^{UV}}$. This second
$t_-(z)$ is interpreted in the AdS-CFT dictionary as a source. All solutions to the original full non-linear equations have to evolve only towards $t_+(z)$
for $z\to0$ in order for the source to vanish. There is however no guarantee that these solutions are finite for $z\to\infty$.
In general it will not be the case, only solutions which evolve to some linear combination $a\,t_+(z)+b\,t_-(z)$ for $z\to0$
will be finite in the opposite limit at $z\to\infty$.
We can enforce $b=0$ and thus have a $t(z)$ sourceless at $z\to0$ and finite at $z\to\infty$
only by carefully choosing the parameters of the original Lagrangian, i.e. the potential.
From here the fine-tuning among parameters.

Another way perhaps more familiar of setting the problem is through
the linearized perturbation equation around the assumed solution $t(z)$.
If we write the perturbation as $\xi(z;q)\,e^{iq\cdot\frac{x}{L}}$,
such equation results (\ref{xieq}).
We can rewrite this linearized equation for perturbations in a Schr\" odinger-like form.
Taking $\,\xi(z;q)=z^\frac{d-1}{2}\,f(z;q)$ we get,
\beq
\label{eforf}
\ddot{f}(z;q)-\left[q^2 + \frac{1}{z^2}\;\left(\frac{d^2-1}{4}+V''(t(z))\right)\right]f(z;q)=0
\eeq
Now, well-known symmetry arguments (in this case related to dilatation invariance)
show that $\xi(z;0)\sim  z\,\dot{t}(z)$ solves equation (\ref{xieq}) with $q^2=0$.
But (\ref{eforf}) is a second order linear differential equation with two independent solutions and then standard quantum mechanics arguments work.
By definition, necessary $f(z;0)\sim  z^{\frac{1}{2} - \nu_{IR}}$ for $z\to\infty$
and the solution that goes as $z^{\frac{1}{2} + \nu_{IR}}$ must be discarded.
Similarly, $f(z;0)\sim  z^{\frac{1}{2} + \nu_{UV}}$ for $z\to 0$
and the solution that goes as $z^{\frac{1}{2} - \nu_{UV}}$ must be discarded too.
The only way for this solution of (\ref{eforf}) to exist is that in both cases we remain with the same function.
As the ``energy" is zero it cannot be quantized as it is usually the case in QM, so $z\,\dot{t}(z)$ can exist only when a fine-tuned relation among parameters in the potential holds,
and so also the solution $t(z)$ of (\ref{eom}) exists only in this case.

\subsection{Fine-tuning the cosmological constant on the boundary}

As it has been explained in \cite{Coradeschi:2013gda} the fine-tuning needed for the potential
parameters is nothing else than the requirement for a vanishing cosmological constant on the boundary.
To see it more explicitly we have of course to reintroduce gravity, i.e. a non-zero $\kappa$.

Let us thus consider the gravity-scalar system defined by the action,
\beq\label{action}
S = \int d^{d+1} x\,\sqrt{|g|}\; \left(\frac{1}{2\,\kappa^2}  \left(  R + \frac{d\,(d-1)}{2} \right)-\frac{1}{2}\,D^M t \; D_M t - V(t)\right)
\eeq
The following equations of motion follow,
\bea\label{eom1}
R_{MN} &=& -d\;g_{MN} + \kappa^2\;\left(T_{MN} - \frac{T^P{}_P}{d-1}\;g_{MN}\right)\cr
D^K D_K t &=& V'(t)
\eea
where the energy momentum-tensor for the scalar field is,
\beq
T_{MN} = D_Mt\; D_Nt - \left(\frac{1}{2}\,D^K t \; D_K t + V(t)\right)\;g_{MN}
\eeq

We are going to consider the ansatz,
\bea\label{ansatz}
g &=& d\rho^2 + A^2(\rho)\;\hat g\cr
t &=& t(\rho)
\eea
where $\hat g \equiv \eta_{mn}\,\hat\omega^m\,\hat\omega^n$ is the metric ($\{\hat\omega^m\}$ is a  vielbein) on a $d$-dimensional space-time with generic coordinates $\Omega$ .
In the obvious local basis,
\beq
\omega^m\equiv A(\rho)\;\hat\omega^m\quad,\quad m=0,1,\dots,d-1\qquad;\qquad \omega^d\equiv d\rho
\eeq
the connections are,
\beq
\omega^m{}_n = \hat\omega^m{}_n\qquad;\qquad \omega^m{}_d= \frac{A'(\rho)}{A(\rho)}\;\omega^m
\eeq
where only in this subsection a prime means $d/d\rho$.

The two-forms defining the curvature tensor result,
\beq
{\cal R}_{mn} = \hat{\cal R}_{mn} - \frac{A'(\rho)^2}{A(\rho)^2}\;\omega_m\wedge\omega_n\qquad;\qquad
{\cal R}_{md} = -\frac{A''(\rho)}{A(\rho)}\;\omega_m\wedge\omega^d
\eeq
Finally the Ricci tensor components are,
\bea\label{ricci}
R_{mn} &=& \frac{1}{A(\rho)^2}\;\hat R_{mn} - \left(\frac{A''(\rho)}{A(\rho)} +
(d-1)\;\frac{A'(\rho)^2}{A(\rho)^2}\right)\;\eta_{mn}\cr
R_{dd} &=& -d\;\frac{A''(\rho)}{A(\rho)}\cr
R_{md} &=& 0
\eea
and the Ricci scalar,
\beq
R = \frac{1}{A(\rho)^2}\;\hat R -2\,d\,\frac{A''(\rho)}{A(\rho)}-
d\,(d-1)\,\frac{A'(\rho)^2}{A(\rho)^2}
\eeq
With (\ref{ansatz}) and (\ref{ricci}) the equations (\ref{eom1}) become,
\bea\label{eom2}
\hat R_{mn} - A^2\left( \frac{A''(\rho)}{A(\rho)} + (d-1)\;\frac{A'(\rho)^2}{A(\rho)^2}
-d + \frac{2\,\kappa^2}{d-1}\;V(t)\right)\;\eta_{mn}&=&0\cr
\frac{A''(\rho)}{A(\rho)} - 1 + \frac{\kappa^2}{d}\;\left(
t'(\rho)^2 + \frac{2}{d-1}\,V(t)\right)&=&0\cr
t''(\rho) + d\,\frac{A'(\rho)}{A(\rho)}\,t'(\rho) - \frac{dV}{dt}(t)&=&0
\eea

We have now two possible cases.
\bigskip

\noindent\underline{Case I:  Vacuum solutions}
\bigskip

Let us consider $t(\rho)=t_v$ an extremum of the potential, $V'(t_v)=0$,
and let us take $V(t_v)=0$. Then there exist three non equivalent, exact
solutions to the gravity equations in (\ref{eom2}),
\bea\label{vacsns}
g &=& d\rho^2 + e^{2\rho}\;\hat g\quad\;\qquad\qquad;\qquad \hat R_{mn} = 0\cr
g &=& d\rho^2 + \cosh^2{(\rho)}\;\hat g\qquad;\qquad \hat R_{mn}
= - (d-1)\,\eta_{mn}\cr
g &=& d\rho^2 + \sinh^2\left(\rho\right)\;\hat g\qquad;\qquad \hat R_{mn}
= + (d-1)\,\eta_{mn}
\eea
We recognize the first case as plane $AdS_{1,d}$ if $\hat g$ is identified with the flat Minkowski metric, the maximally symmetric case.
On the other hand, the second/third solutions correspond to Einstein space-times
of negative/positive curvature, being the most symmetric choices for $\hat g$ the spaces
$AdS_{1,d-1}/dS_{1,d-1}$ with scale $L=1$.
However from (\ref{eom1}) we see that in any case the equation for the bulk metric $g$ is
just $R_{MN} = -d\;g_{MN}$; so the maximally symmetric choices should lead to
the same space, i.e. the three cases in (\ref{vacsns}) must correspond to $AdS_{1,d}$ sliced differently.
\footnote{
In fact the third form can be found in equation (3.1) of  \cite{Maldacena:2010un}. 
}
An observation: $z\equiv e^{-\rho}$ is the usual coordinate with $z=0$ the
boundary and $z=\infty$ the horizon iff the $\rho$-coordinate is the one defined in the patch of the first solution, i.e. $\rho$ represent different coordinates in each line of (\ref{vacsns}).
\bigskip

\noindent\underline{Case II:  Domain wall solutions}
\bigskip

In this case the profile of the scalar must be non trivial; in particular we are interested
in interpolating solutions like the ones considered in the papers.
We can however always take the weak gravity, decoupling limit $\kappa\rightarrow 0$,
and we must solve the scalar equation in the background  (\ref{vacsns}).
Now, if we consider the flat slicing, we found the need of fine-tuning the potential in order to get
a solution.
The question is: if we interpret the other two slicings as leading to $AdS$ and $dS$ space-time geometries
of the boundary theory instead of Minkowski, is it necessary to fine-tuning the potential to get a
domain wall solution also in these cases?

With the new variable $z=e^{-\rho}$ (and for simplicity keeping the same notation for $t=t(z)$)
the equation to solve is,
\beq
z^2\ddot t(z) - \frac{(d-1)+k(d+1)z^2}{1-kz^2}z\dot t(z) - \frac{dV}{dt}(t) = 0
\eeq
where $k=+1, 0, -1$ in the $dS$, Minkowski, $AdS$ slicing. There is no dilatation symmetry anymore, so no need
for fine-tuning. In the language of the piece-wise-quadratic potential, all the coordinates of different intervals can be
determined, and not only ratios. No relations among parameters is needed for the solution to exist. It is now clear the
physical meaning of it: it is just the fine-tuning of the boundary cosmological constant.

\subsection{The BPS solutions}

A solution that spontaneously breaks conformal invariance makes the on-shell action vanish (see for example
\cite{Berman:2002kd}). This is a hint that the solution may be of the BPS type, i.e. it solves a first order equation
\cite{Bajc:2012vk}. Instead of proving this statement, we will show how one can define the superpotential that
allows a smooth $\kappa\to0$ limit. Let's go back to (\ref{action}). We will search for solutions to the equations
of motion of the form
\bea
g &=& \frac{1}{z^2}\left( d{\vec x}^2+ L^2\;\frac{dz^2}{F(z)}\right)\cr
t &=& t(z)
\eea
The b.c. at the boundary $z=0$ are,
\beq\label{bcbound}
t(z)\rightarrow 0\qquad;\qquad F(z)\rightarrow 1
\eeq
where,
\beq
V(0) = 0\qquad;\qquad V'(t)|_{t=0} = 0
\eeq
This assures for the solution to be asymptotically AdS with fixed radius $L=1$.

At the horizon $z=\infty$ we impose,
\beq\label{bchor}
t(\infty)<\infty\quad;\quad F(z) = F_h + {\cal O}\left(\frac{1}{z}\right)
\eeq

The equations of motion result
\bea\label{fullem2}
z\,F'(z) &=& \kappa^2\;z^2\; F(z)\; \dot t^2(z)\cr
z^{d+1}\; F^\frac{1}{2}(z) \frac{d}{dz}\left(\frac{F^\frac{1}{2}(z)}{z^{d-1}}\; \frac{dt}{dz}(z)\right)&=& V'(t)
\eea
With no back-reaction ($\kappa=0$), $F(z)=1$ and it is the second equation to solve, just the scalar
fields in the AdS background. When back-reaction is taken into account ($\kappa> 0$) we can use the superpotential trick.
The usual choice is consider potentials which can be written as\beq
V(t) = \frac{1}{2}\, W'^2(t)-\frac{d\,\kappa^2}{4}\,W^2(t) + \frac{d}{\kappa^2}
\eeq
Then it is possible to show that a solution of,
\bea
F(z)&=& \frac{\kappa^4}{4}\;\left.W^2(t)\right|_{t= t(z)}\cr
t'(z) &=& \frac{2}{\kappa^2\,z}\;\left.\frac{W'(t)}{W(t)}\right|_{t= t(z)}
\eea
solves (\ref{fullem2}).

This ansatz implicitly assume that $\kappa\ne0$. On the other side, if we want eventually to get
the no-backreaction limit $\kappa\to0$, we choose a potential of the form
\beq
V(t) = \frac{1}{2}\, W'^2(t)\ - d\,W(t)- \frac{\kappa^2\,d}{4}\;W^2(t)
\eeq
It is then possible to show that a solution of,
\bea
F(z)&=& \left.H^2(W(t))\right|_{t= t(z)}\cr
z\;\dot t(z) &=& \left.\frac{W'(t)}{H(W(t))}\right|_{t= t(z)}
\eea
where
\beq
H(W) \equiv 1 + \frac{\kappa^2}{2}\; W
\eeq
is a solution of (\ref{fullem2}). The $\kappa\to0$ limit is now small and points toward
the potential
\beq
\label{Vatkappa0}
V(t) = \frac{1}{2}\, W'^2(t)\ - d\,W(t)
\eeq
and the following BPS like equation,
\beq
\label{eomBPS}
z\;\dot{t}(z)= W'(t(z))
\eeq
whose solutions satisfy also the full second order equation of motion (\ref{eom})
and for which the action (\ref{action}) vanishes.

At least for polynomial superpotentials and potentials the fine-tuning for vanishing boundary cosmological
constant is simply the special form (\ref{Vatkappa0}). With this we mean that all coefficients of the
polynomial in the potential are not independent and thus the potential itself is not generic.

Before ending this section, let us see some examples of superpotentials $W(t)$
(in \cite{Bajc:2013wha} we already showed another choice).

\subsubsection{\label{Z2}The $Z_2$ symmetric case}

An interesting case consists of the sixth order potential with the $Z_2$ symmetry $t\to -t$.
The ansatz for the superpotential
\beq
W(t)=\Delta\;\left(\frac{1}{2}\;t^2 - \frac{1}{4}\;t^4\right)
\eeq
leads to the solution,
\beq
t(z)=\frac{z^\Delta}{\left(1+z^{2\Delta}\right)^{1/2}}
\eeq
From here we see that
\beq
\Delta^{UV}=\Delta\;\;\;,\;\;\;\Delta^{IR}= d + 2\,\Delta
\eeq

\subsubsection{\label{UVIR}A case with $\Delta^{UV}$ and $\Delta^{IR}$ independent}

In the examples of \cite{Bajc:2013wha} and above a correlation between the UV and IR $\Delta$'s was present.
This is however not a generic feature of the system.
In fact, choosing for example
\beq
W(t)=-\frac{1}{4}\;\left(\Delta^{IR}-(d+\Delta^{UV})\right)\;t^4+
\frac{1}{3}\;\left(\Delta^{IR}-(d+\Delta^{UV})-\Delta^{UV}\right)\;t^3
+\frac{\Delta^{UV}}{2}\;t^2
\eeq
we get the solution
\beq
z(t)=\left[\frac{\Delta^{UV}+\left(\Delta^{IR}-(d+\Delta^{UV})\right)\,t}{1-t}\right]^{\frac{1}
{\Delta^{IR}-d}}
\left[\frac{t}{\Delta^{UV}+\left(\Delta^{IR}-(d+\Delta^{UV})\right)\,t}\right]^{\frac{1}{\Delta^{UV}}}
\eeq
which has the limits (\ref{bc}) with
\beq
a_{UV} = \left(\Delta^{UV}\right)^{\frac{\Delta^{IR}-\left(d+\Delta^{UV}\right)}{\Delta^{IR}-d}}\qquad;\qquad
a_{IR} = -\left(\Delta^{IR}-d\right)^{-\frac{\Delta^{IR}-\left(d+\Delta^{UV}\right)}{\Delta^{UV}}}
\eeq
The parameters $\Delta^{UV}>d/2$ (corresponding to the maximum or minimum in the UV) and
$\Delta^{IR}>d+\Delta^{UV}$ (minimum in the IR) can be otherwise arbitrary.

\section{\label{mass}${\bf V''(t)<-d^2/4}$}

As we said before, in a piece-wise quadratic potential at least in some interval the second derivative of
the potential must be smaller than $-d^2/4$ for the solution to exist. Let us here show this statement for
a general potential $V(t)$ characterized by (\ref{Vatkappa0}). Let us define
\beq
\label{F0}
\left.F\equiv\int d\mu\;W'(t)^2\;\left(V''(t)+\frac{d^2}{4}\right)\right|_{t=t(z)}
\eeq
where $t(z)$ is the solution of the BPS equation (\ref{eomBPS}) and to simplify the
notation we will use in this subsection the abbreviation
\beq
\int d\mu\dots\equiv\int_0^\infty dz\; z^{-d-1}\dots
\eeq
and omit the field dependence.
Our aim is to show that the quantity $F$ is non-positive, so that $V''<-d^2/4$ at least
in some region.

First we rewrite (\ref{F0}) using (\ref{Vatkappa0})
\beq
\label{F1}
F
=\int d\mu\,\left(W'^2\;W''^2+W'^3\;W'''-d\;W'^2\;W''+\frac{d^2}{4}\;W'^2\right)
\eeq
Now we use (assuming vanishing boundary terms, which is easily verified)
\bea
\label{rel1}
\int d\mu \;W'^2&=&
\frac{2}{d}\int d\mu\; W'^2\;W''\\
\label{rel2}
\int d\mu\; W'^3\;W'''&=&
\int d\mu\left(d\; W'^2\;W''-2\;W'^2\;W''^2\right)
\eea
to rewrite (\ref{F1}) as
\beq
F=\frac{d}{2}\int d\mu\; W'^2\;W''-\int d\mu\; W'^2\;W''^2
\eeq
Finally we use the Schwartz inequality
\beq
\label{schwartz}
\int d\mu\; f\;g\leq\left(\int d\mu\; f^2\right)^\frac{1}{2}\;
\left(\int d\mu\; g^2\right)^\frac{1}{2}
\eeq
to derive from (\ref{rel1})
\beq
\int d\mu\;W'^2\leq\frac{4}{d^2}\int d\mu\; W'^2\;W''^2
\eeq
Using then (\ref{schwartz}) we get first
\beq
\int d\mu\; W'^2\;W''\leq\left(\int d\mu\; W'^2\right)^\frac{1}{2}\;
\left(\int d\mu\; W'^2\;W''^2\right)^\frac{1}{2}
\eeq
from which finally it follows
\beq
F\leq 0
\eeq
This proves our statement: the inequality $V''<-d^2/4$ is valid at least in some region of $z$
for any potential $V$ of the form (\ref{Vatkappa0}).

Notice that since at the horizon ($z\to\infty$) the potential has a minimum and at the
boundary ($z=0$) a minimum or a maximum in the conformal window (i.e. $V''+d^2/4>0$),
there are always an even number of times that $V''$ crosses the particular value $-d^2/4$.

\section{\label{WKB}The WKB approximation method}

Here we shall try to apply the WKB method in order to compute the two-point correlation function
of operators dual through the AdS/CFT correspondence to a bulk scalar field.
The recipe to get it is to consider the solution to the perturbation equation (\ref{xieq}), and identify  the propagator by looking at the behavior near the boundary $z\rightarrow0$,
\beq\label{propag}
\xi(z;q)\sim z^{d-\Delta^{UV}} + G_2(q)\; z^{\Delta^{UV}}
\eeq

The straightest way of doing it is to consider the Schr\"odinger-type equation
(\ref{eforf}) with ``potential"
\beq\label{Q}
Q(z;q)\equiv q^2 + \frac{1}{z^2}\;\left( \frac{d^2-1}{4} + V''(t(z))\right)
\eeq
where we remember that $t(z)$ is the solution of (\ref{eom}).
For simplicity we consider the case $V''(0)\equiv m^2_{UV}>0$,
although it is not necessary for the argument.

The WKB approximation results a good one if the slowly varying ``Compton length" condition holds,
\beq\label{valeWKB}
\left|\frac{d|Q(z;q)|^{-\frac{1}{2}}}{dz}\right| = \left|\frac{\dot Q(z;q)}{2\,|Q(z;q)|^\frac{3}{2} }\right|\ll 1
\eeq
This condition applied to (\ref{Q}) reads,
\beq\label{valeWKBour}
\frac{\left| \frac{d^2-1}{4} + V''(t(z)) - \frac{1}{2}\,V'''(t(z))\,z\,\dot t(z)\right|}
{\left|\frac{d^2-1}{4} + V''(t(z)) + q^2\,z^2\right|^\frac{3}{2}}\ll 1
\eeq
From here is straightforward to see that the WKB solution is trustable for any $q^2$ around $z=0$ and $z=\infty$ if,
\beq
\nu_{UV}\gg\frac{1}{2}\qquad;\qquad \nu_{IR}\gg\frac{1}{2}
\eeq
respectively, with $\nu_{UV/IR}$ as in (\ref{Delta}).
Furthermore, $Q(z;q)$ is positive near $z=0$ (and diverges quadratically there), but it is also positive for large $z$ (going to $q^2$ from above).
What happens in the middle?
From section $5$ we know that for $q$ small enough $Q(z;q)$ must become negative;
then for some $z_M$ where $t(z_M)= t_M$ it should have a local minimum.
Then there must exist $z_i=z_i(q), \,z_1(q)<z_M<z_2(q)$ such that,
\beq
z_i{}^2\; Q(z_i;q) = \frac{d^2-1}{4} + V''(t(z_i)) + q^2\, z_i{}^2 = 0\quad; \quad i=1,2
\eeq
Near these zeroes of $Q(z;q)$ the WKB approximation breaks down.

If we admit that $V''(t_M)\;$ is large enough then it is seen from (\ref{valeWKBour})
that in the region near $z_M$ the WKB solution is trustable too.
Therefore, calling $I, II,III$ the regions near $z=0,\, z_M$ and $z\gg 1$ respectively,
we can write the approximate WKB solution in each region as,
\bea
\xi_I(z;q) &=&C_I^+\;z^\frac{d}{2}\;\frac{\exp\left(\int_{z_1}^z\,\frac{dz}{z}\,\sqrt{z^2\,Q(z;q)}
\right)}{(z^2\,Q(z;q))^\frac{1}{4}}+C_I^-\;z^\frac{d}{2}\;\frac{\exp\left(-\int_{z_1}^z\,\frac{dz}{z}\,
\sqrt{z^2\,Q(z;q)}\right)}{(z^2\,Q(z;q))^\frac{1}{4}}\cr
&&\\
\xi_{II}(z;q) &=&
C_{II}\;z^\frac{d}{2}\;\frac{\exp\left(i\int_{z_1}^z\,\frac{dz}{z}\,\sqrt{-z^2 Q(z;q)}\right)}
{\left(-z^2\,Q(z;q)\right)^\frac{1}{4}}+
C_{II}^*\;z^\frac{d}{2}\;\frac{\exp\left(-i\int_{z_1}^z\,\frac{dz}{z}\,\sqrt{-z^2Q(z;q)}\right)}
{\left(-z^2\,Q(z;q)\right)^\frac{1}{4}}\cr
&&\\
\xi_{III}(z;q) &=&
C_{III}^+\;z^\frac{d}{2}\;\frac{\exp\left(\int_{z_2}^z\,\frac{dz}{z}\,\sqrt{z^2\,Q(z;q)}\right)}
{(z^2\,Q(z;q))^\frac{1}{4}}
+ C_{III}^-\;z^\frac{d}{2}\;\frac{\exp\left(-\int_{z_2}^z\,\frac{dz}{z}\,\sqrt{(z^2\,Q(z;q)}\right)}
{(z^2\,Q(z;q))^\frac{1}{4}}\cr
& &
\eea
where the coefficients are related by,
\bea\label{coeffrel1}
C_I^\pm &=& \frac{1\pm3}{2}\;Im\left(C_{II}\;e^{\pm i\frac{\pi}{4}}\right)
\quad\leftrightarrow\quad
C_{II} =\frac{1}{2}\;e^{+i\frac{\pi}{4}}\;C_I^+ + e^{-i\frac{\pi}{4}}\;C_I^- = (C_{II}^*)^*\\
\label{coeffrel2}
C_{III}^\pm &=& \frac{3\pm1}{2}\;Im\left(C_{II}\;e^{i(\varphi(q)\mp \frac{\pi}{4})}\right)
\quad\leftrightarrow\quad\cr
C_{II} &=& e^{-i\varphi(q)}\left(-\frac{1}{2}\;e^{-i\frac{\pi}{4}}\;C_{III}^+
+ e^{+i\frac{\pi}{4}}\;C_{III}^-\right) = (C_{II}^*)^*
\eea
and,
\beq
\varphi(q) \equiv \int_{z_1(q)}^{z_2(q)}\,\frac{dz}{z}\,\sqrt{-z^2\,Q(z;q)}
\eeq

Now, imposing finiteness when $z\rightarrow \infty$ implies $C_{III}^+ =0$.
By using the relations (\ref{coeffrel1}) and (\ref{coeffrel2}) we get all the
constants in terms of $C_{III}^-$; in particular for the solution near $z=0$ we get,
\bea
\label{xiI}
\xi_I(z;q) &=& C_{III}^-\,\left( 2\,\cos\varphi(q)\;z^\frac{d}{2}\;
\frac{\exp\left(\int_{z_1(q)}^z\,\frac{dz}{z}\,\sqrt{z^2\,Q(z;q)}
\right)}{(z^2\,Q(z;q))^\frac{1}{4}}\right.\cr
&+&\left.\sin\varphi(q)\;z^\frac{d}{2}\;\frac{\exp\left(-\int_{z_1(q)}^z\,\frac{dz}{z}\,
\sqrt{z^2\,Q(z;q)}\right)}{(z^2\,Q(z;q))^\frac{1}{4}}\right)
\eea
From here we should be able to extract the propagator as a function of $q^2$,
at least for $q$ not so large.
But we know from section $4$ that for $q=0$ (\ref{xiI}) must be equal to
$z\,t'(z)$ and thus going only as $z^{\Delta^{UV}}$ for $z\to 0$.
We will show now that this implies the constraint $\varphi(0)=k\,\pi$ with $k$ an integer.
First we rewrite
\bea
\label{expz1}
\exp{\left(\pm\int_{z_1(q)}^z\frac{dz}{z}\sqrt{z^2\,Q(z;q)}\right)}&=&
\left(\frac{z}{z_1(q)}\right)^{\pm\sqrt{\nu_{UV}^2-1/4}}\\
&\times&\exp{\left(\pm\int_{z_1(q)}^z\frac{dz}{z}\left(\sqrt{z^2\,Q(z;q)}-\sqrt{\nu_{UV}^2-1/4}\right)\right)} \nonumber
\eea
Since we are interested only in $\nu_{UV}\gg 1/2$ and leading behavior at $z\to 0$, we
can see with the help of (\ref{expz1}) that the first term on the r.h.s. of (\ref{xiI}) goes like $z^{\Delta^{UV}}$, while the second goes like $z^{d-\Delta^{UV}}$.
Since this last one should not be present in the solution $z\,t'(z)$ of the $q=0$ perturbation, we have to impose (otherwise no solution with the right asymptotic behavior exists)
\beq
\label{phi0}
\varphi(0)\equiv\int_{z_1(0)}^{z_2(0)}\frac{dz}{z}\sqrt{-z^2\,Q(z;0)}=k\;\pi
\eeq
This means that only potentials which satisfy this constraint are acceptable.
This is the WKB analog of the fine-tuning mentioned before.

This simple conclusion is the reason for the $1/q^2$ behavior of the boundary propagator. In fact, it is easy to derive the form of the propagator in the WKB approximation; from (\ref{propag}) we get:
\beq
G_2(q)=\frac{2\exp{\left(-2\int_0^{z_1(q)}\frac{dz}{z}\left(\sqrt{z^2\,Q(z;q)}-
\sqrt{\nu_{UV}^2-1/4}\right)\right)}} {(z_1(q))^{2\sqrt{\nu_{UV}^2-1/4}}\tan{\varphi(q)}}
\eeq
Clearly, due to (\ref{phi0}), we get for $q\to 0$ the usual Goldstone pole
\beq
G_2(q)\approx \frac{2\exp{\left(-2\int_0^{z_1(0)}\frac{dz}{z}\left(\sqrt{z^2Q(z;0)}-\sqrt{\nu_{UV}^2-1/4}\right)\right)}}
{(z_1(0))^{2\sqrt{\nu_{UV}^2-1/4}}(d\varphi(q)/dq^2)_{q^2=0}}\times\frac{1}{q^2}
\eeq
where
\beq
\left.\frac{d\varphi(q)}{dq^2}\right|_{q^2=0}=
-\frac{1}{2}\int_{z_1(0)}^{z_2(0)}dz\frac{z}{\sqrt{-z^2Q(z;0)}}
\eeq
Although the denominator vanishes at the integration boundaries, the integral itself is finite.







\section{\label{allorders}The matching method to all orders}
\cleqn

Let $t(z)$ be the solution of the equation of motion (\ref{eom})
that behaves for $z\rightarrow 0$ (UV) and $z\rightarrow \infty$ (IR) as,
\beq\label{asympt}
t(z) \xrightarrow{z\rightarrow 0} a_{UV}\;z^{\Delta^{UV}_+}\;(1+ b_{UV}\,z^{\alpha_{UV}}+\dots)\qquad;\qquad
t(z) \xrightarrow{z\rightarrow\infty} 1 + a_{IR}\;z^{\Delta^{IR}_-}\;(1+ b_{IR}\,z^{\alpha_{IR}}+\dots)
\eeq
respectively.
Here $\alpha_{UV}>0$ and $\alpha_{IR}<0$, while that
\beq
\Delta^{UV/IR}_\pm \equiv \frac{d}{2} \pm \nu_{UV/IR}\qquad;\qquad
\nu_{UV/IR}\equiv\sqrt{\frac{d^2}{4} + m^2_{UV/IR}}\
\eeq
with $m^2_{UV}\equiv V''(0)$ and $m^2_{IR}\equiv V''(1)>0$.
Note that in order for $t(z)$ to be finite in the asymptotic expansions (\ref{asympt})
neither $\Delta^{UV}_-$ appears in the UV nor $\Delta^{IR}_+$ in the IR.

Let us introduce for further use the following expansions of the functions $\xi_\pm(z)$
\beq\label{asymxi+-}
\xi_\pm(z) = a_\pm^{UV/IR}\;z^{\Delta_{\pm/\mp}^{UV/IR}}\;\bar\xi_\pm^{UV/IR}(z)
\qquad;\qquad\bar\xi_\pm^{UV/IR}(z)\xrightarrow{z\rightarrow 0/\infty} 1
\eeq
that follow by plugging (\ref{asympt}) in the definitions
\bea
\xi_+(z)&\equiv&\; z\,\dot{t}(z)\label{xi+}\\
\xi_-(z)&\equiv&\xi_+(z)\left(\int_{z_i}^z dy\frac{y^{d-1}}{\xi_+^2(y)}+\frac{\xi_-(z_i)}{\xi_+(z_i)}\right)
\label{xi-}
\eea
where $z_i$ and $\xi_-(z_i)$ are integration constants. We find
\beq
a_+^{UV/IR} \equiv a_{UV/IR}\;\Delta^{UV/IR}_{+/-}\qquad;\qquad
a_-^{UV/IR} \equiv \left(a_+^{UV/IR}\;(d-2\,\Delta^{UV/IR}_{+/-})\right)^{-1}
\eeq
Clearly the UV/IR expansion of $\xi_+(z)$ can not contain the $z^{\Delta^{UV/IR}_{-/+}}$-power, but $\xi_-(z)$ could contain the $z^{\Delta^{UV/IR}_{+/-}}$-power.

Our aim is to solve the equation for perturbations around the solution $t(z)$, i.e.
if we write (for a general treatment see for example the appendix of \cite{Bajc:2012vk})
\beq
t(z;q) \equiv t(z) + \xi(z;q)\; e^{iq\cdot\frac{x}{L}}
\eeq
then the second equation in (\ref{eom1})  gives to first order in $\xi(z;q)$
\beq
\label{xieq}
z^2\;\ddot{\xi}(z;q)-(d-1)\;z\;\dot{\xi}(z;q)-\left(q^2\,z^2 + V''(t(z))\right)\;\xi(z;q)=0
\eeq
We will do it in two different approximations.


\subsection{\label{largez}The large $z$ expansion.}

We write (\ref{xieq}) as
\beq
\label{xieqLz}
z^2\;\ddot{\xi}(z;q)-(d-1)\;z\;\dot{\xi}(z;q)-\left(q^2\,z^2 + m^2_{IR} \right)\;\xi(z;q)=
\delta(z)\;\xi(z;q)
\eeq
and consider $\delta(z)\equiv V''(t(z))-V''(1)$ small in the sense,
\beq\label{delta}
|\delta(z)| = |V''(t(z))-V''(1))|\ll V''(1) \quad\longrightarrow\quad
z>z_\infty\equiv\left|\frac{V''(1)}{V'''(1)\;a_{IR}}\right|^\frac{1}{\Delta^{IR}}
\eeq
independently of the value of $q$.
Then the solution for $z>z_\infty$ can be hopefully expanded in orders of
$\delta(z)$,
\beq
\delta(z) = V'''(1)\;a_{IR}\;z^{\Delta^{IR}}\;\left(1 + b_{IR}\,z^{\alpha_{IR}}+\dots\right)
\eeq
The order zero term is the solution to the l.h.s. of (\ref{xieqLz}) equal to zero,
which is given by,
\beq\label{xiinfty}
\xi_\infty(z;q) \equiv \frac{2}{\Gamma(\nu_{IR})}\;
\left(\frac{q}{2}\right)^{\nu_{IR}}\;z^\frac{d}{2}\;K_{\nu_{IR}}(q z)
\eeq
where we have dropped the solution that diverges in the IR and fixed the normalization
in such a way that $\xi_\infty(z;0) = z^{\Delta^{IR}_-}$.
It is not difficult to see that the expansion for large $z>z_\infty$ is of the form,
\beq\label{xiexp0}
\xi(z;q) = \xi_\infty(z;q)\;\left( 1 + \frac{f_0(q z)}{z^{-\Delta_-^{IR}}} + \dots\right)
\eeq
where for completeness we quote the first correction,
\beq
f_0(u) = V'''(1)\,a_{IR}\;x^{-\Delta_-^{IR}}\;\int_{\infty}^u \frac{dx}{x\,K^2_{\nu_{IR}}(x)}\int_{\infty}^x \frac{dy}{y^{1-\Delta_-^{IR}}} \,K^2_{\nu_{IR}}(y)
\eeq
However corrections to the leading term of $\xi(z;q)$ in negative powers of $z$ will not be relevant in the matching procedure, at least not to compute the leading order behavior of the two-point function.



\subsection{The small $q$ expansion.}

This time we write (\ref{xieq}) as
\beq
\label{xieqSqz}
z^{d-1}\;\frac{d}{dz}\left(z^{1-d}\frac{d\xi(z;q)}{dz}\right)-\frac{V''(t(z))}{z^2}\;\xi(z;q)
= q^2\;\xi(z;q)
\eeq
and consider $q$ small in the sense,
\beq
q\ll \frac{|V''(t(z))|^\frac{1}{2}}{z}
\eeq
This condition certainly holds in the UV region near $z=0$ , but also in the IR region if
\beq\label{IRval}
q^2\,z^2 \ll |V''(t(z)|\sim m^2_{IR} \quad\longrightarrow\quad q z\ll m_{IR}
\eeq
that is, when $z$ is large and q small but $q z$ fixed and small enough.

Under this condition we can try a solution for small $q$ as a power series in $q^2$,
\beq\label{xiexp1}
\xi(z;q) = \sum_{m\geq 0}q^{2m}\;\xi^{(m)}(z;q)
\eeq
Plugging this expansion in (\ref{xieqSqz}) we get,
\bea
\label{xi0eq}
z^{d-1}\;\frac{d}{dz}\left(z^{1-d}\,\frac{d\xi^{(0)}(z;q)}{dz}\right)
-\frac{V''(t(z))}{z^2}\;\xi^{(0)}(z;q)&=& 0\\
\label{ximeq}
z^{d-1}\;\frac{d}{dz}\left(z^{1-d}\,\frac{d\xi^{(m)}(z;q)}{dz}\right)
-\frac{V''(t(z))}{z^2}\;\xi^{(m)}(z;q)
&=& \xi^{(m-1)}(z;q)\quad;\quad m=1,2,\dots\nonumber\\
\eea
The solution to lowest order is,
\beq\label{xi0}
\xi^{(0)}(z;q) = C^{(0)}_+(q)\;\xi_+(z) + C^{(0)}_-(q)\;\xi_-(z)
\eeq
where $C^{(0)}_\pm(q)$ are integration constants.
With  $\xi^{(0)}(z;q)$ we can determine $\xi^{(1)}(z;q)$ from (\ref{ximeq}), and so on.

This iterative procedure yields the solution in the following form.
First we introduce the set of functions,
\bea\label{fij}
f_{ij}^{(k)}(z) &\equiv& \int_{z_i}^z\,\frac{d w}{w^{d-1}}\;\xi_i(w)\;\xi_j^{(k)}(w)
\quad;\quad i,j=+,-\quad,\quad k=0,1,\dots\cr
\xi_\pm^{(0)}(z) &\equiv& \xi_\pm(z)
\eea
where
\beq\label{xi(k)}
\xi_\pm^{(k)}(z) \equiv -f_{-\pm}^{(k-1)}(z)\;\xi_+(z)
+ f_{+\pm}^{(k-1)}(z)\;\xi_-(z)\quad,\quad k=1,2,\dots
\eeq
All of them are obtained iteratively: first, from (\ref{fij}) with $k=0$
we get $f_{ij}^{(0)}(z)$, then we go to (\ref{xi(k)}) with $k=1$ and get $\xi_\pm^{(1)}(z)$,
then we come back to (\ref{fij}) with $k=1$ and get $f_{ij}^{(1)}(z)$ and so on.
The functions $\xi_m(z;q)$ can be expressed in terms of the $\xi_\pm^{(k)}(z)$'s yielding
the full expansion (\ref{xiexp1}) in the form,
\beq\label{xiexp2}
\xi(z;q) = \sum_{m\geq 0}q^{2m}\;\sum_{k=0}^m\;\left(
C^{(m-k)}_+(q)\;\xi_+^{(k)}(z) + C^{(m-k)}_-(q)\;\xi_-^{(k)}(z)\right)
\eeq
where the  $C^{(k)}_\pm$'s are, as in (\ref{xi0}), the integration constants of the
homogeneous solution in (\ref{ximeq}).
After some rearrangement, we can write (\ref{xiexp2}) as,
\beq\label{xiexp3}
\xi(z;q) = C_+(q)\;\sum_{m\geq 0}\;q^{2m}\;\xi_+^{(m)}(z) +
C_-(q)\;\sum_{m\geq 0}\;q^{2m}\;\xi_-^{(m)}(z)
\eeq
where we have redefined the coefficients
\beq
C_\pm(q)\equiv\sum_{k\geq 0}\;C^{(k)}_\pm(q)\;q^{2k}
\eeq
We should not be surprised of this expression; after all (\ref{xieqSqz}) is a second order linear differential equation and both sums in (\ref{xiexp3}) are linearly independent solutions of it as it can be quickly checked.
Note furthermore that they are holomorphic in $q^2$; the reason behind this fact can
be traced directly to the assumption (\ref{xiexp1}).


\subsection{\label{appA}The two-point function.}

For $q z\ll  m_{IR}$ expansion (\ref{xiexp3})
hopefully holds, and it can be used to compute the two-point correlation function at low momenta as follows.
After adjusting the constant of integration in (\ref{asympt}) to get rid of the
$z^{\Delta^{UV}_+}$ term in $\xi_-(z)$, we parametrize the $z\to0$ behavior as
\bea
\xi(z;q)&\to&\left[\left(1-q^2\epsilon_{-+}^{UV}(q)\right)C_+(q)-
q^2\epsilon_{--}^{UV}(q)C_-(q)\right]a_+^{UV}z^{\Delta_+^{UV}}+\ldots\\
&+&\left[\left(1+q^2\epsilon_{+-}^{UV}(q)\right)C_-(q)
+q^2\epsilon_{++}^{UV}(q)C_+(q)\right]a_-^{UV}z^{\Delta_-^{UV}}+\ldots\nonumber
\eea
where
\beq
\label{epsilonppm}
\epsilon_{+\pm}^{UV}(q)=-\sum_{m\ge 0}q^{2m}\int_0^{z_i}dw\;w^{1-d}\; \xi_+(w)\;\xi_\pm^{(m)}(w)
\eeq
while we were unable to find a closed expression for $\epsilon_{-\pm}^{UV}$ without specifying the potential.

Applying the holographic recipe (\ref{propag}) the two-point function results,
\beq\label{G21}
G_2(q)\xrightarrow{q\rightarrow 0} \frac{a_+^{UV}/a_-^{UV}}{\left(q^2\;\epsilon_{++}^{UV}(q) +
\frac{C_-(q)}{C_+(q)}\right)_{q\rightarrow 0}}
\eeq

The knowledge of the leading order behavior of the quotient
$C_-(q)/C_+(q)$ for $q\rightarrow 0$ will allow to compute
the leading power in $q$ of $G_2(q)$.
The $z_i$-dependence of the coefficients $\epsilon_{++}^{UV}$ (and the $z_i$-independence of the
physics) gives a hint that this power is $-2$, as we will confirm below.


\subsection{The infrared expansion}

Here we define the functions $\bar F_{ij}^{(m)}(z)$
and the constants $\bar\varphi_{ij}^{(m)}$ by means of the integrals,
\bea\label{Fijm}
& &a_i^{IR}\,a_j^{IR}\,\sigma^{(m)}_j\;\int_{z_i}^z\,d w\;
w^{1+2m+\Delta_{(i)} +\Delta_{(j)}-d}\;\bar\xi^{(0)}_i(w)\;\bar\xi^{(m)}_j(w)\cr
&\equiv& \bar\varphi_{ij}^{(m)} + \frac{a_i^{IR}\,a_j^{IR}\,\sigma^{(m)}_j\;
z^{2+2m+\Delta_{(i)} +\Delta_{(j)}-d}}{2+2m+\Delta_{(i)} +\Delta_{(j)}-d}
\bar F_{ij}^{(m)}(z)\qquad;\quad m=0,1,\dots
\eea
where $\bar\varphi_{ij}^{(m)}$ is defined to be the only $z$-independent part in the large $z$ expansion,
and
\beq\label{sigma}
\sigma^{(m)}_\pm \equiv \frac{\Gamma(1\mp\nu_{IR})}{2^{2m}\,m!\,\Gamma(1\mp\nu_{IR}+m)}
\qquad;\qquad m=0,1,\dots
\eeq
With them we can calculate ($m=1,2,\dots$),
\bea\label{barxim}
\varphi_{ij}^{(m)}&\equiv&\bar\varphi_{ij}^{(m)}+
\sum_{k=0}^{m-1}\left(\bar\varphi_{i-}^{(k)}\;\varphi_{+j}^{(m-1-k)}-
\bar\varphi_{i+}^{(k)}\;\varphi_{-j}^{(m-1-k)}\right)\\
\bar\xi_\pm^{(m)}(z)&\equiv& \frac{1}{\nu_{IR}}\;\left(
(\nu_{IR}\mp m)\;\bar F_{\mp\pm}^{(m-1)}(z)\;\bar\xi_\pm^{(0)}(z)\pm
m\;\bar F_{\pm\pm}^{(m-1)}(z)\;\bar\xi_\mp^{(0)}(z)
\right)
\eea
The general form of $\xi_\pm^{(m)}(z)$ for $m=1,2,\dots$, results,
\bea\label{xim}
\xi_\pm^{(m)}(z) &=& a_\pm^{IR}\;\sigma^{(m)}_\pm\;z^{\Delta_\mp^{IR}+2m}\;\bar\xi_\pm^{(m)}(z)\\
&+&\sum_{k=0}^{m-1}\left(
-a_+^{IR}\;\sigma^{(k)}_+\;\varphi_{-\pm}^{(m-1-k)}\;z^{\Delta_-^{IR}+2k}\;\bar\xi_+^{(k)}(z)
+a_-^{IR}\;\sigma^{(k)}_-\;\varphi_{+\pm}^{(m-1-k)}\;z^{\Delta_+^{IR}+2k}\;\bar\xi_-^{(k)}(z)
\right)\nonumber
\eea
where the ingredients to construct it are iteratively computed as described above.

\subsection{\label{appmatching}The matching procedure.}

According to (\ref{delta}) and (\ref{IRval}), in the region
\beq
z>z_\infty\qquad;\qquad
x\equiv q z\ll m_{IR}
\eeq
both expansions (\ref{xiexp0})
and (\ref{xiexp3}) hold and therefore they should coincide {\it exactly}, i.e.
\beq\label{matcheq}
\xi_\infty(z;q)\;\left( 1 + \frac{f_0(q z)}{z^{-\Delta_-^{IR}}} + \dots\right)
= C_+(q)\;\sum_{m\geq 0}\;q^{2m}\;\xi_+^{(m)}(z) +
C_-(q)\;\sum_{m\geq 0}\;q^{2m}\;\xi_-^{(m)}(z)
\eeq
This equation must be used to compute the unknown coefficients $C^\pm(q)$.
As we will see shortly, this is not an easy task in general; fortunately the leading
order behavior necessary to compute (\ref{G21}) is relatively simple to get.
To proceed we need the IR behavior of the $\xi_\pm^{(m)}(z)$'s.
By plugging (\ref{xim}) in (\ref{xiexp3}) we get,
\bea\label{xiexpIR}
&&z^{-\Delta_-^{IR}}\;\xi(z;q) = z^{-\Delta_-^{IR}}\;r.h.s.\,{(\ref{matcheq})}\cr &=&
\left(\left(1-q^2\;\epsilon_{-+}^{IR}(q)\right)\,C_+(q)
- q^2\;\epsilon_{--}^{IR}(q)\,C_-(q)\right)\,a_+^{IR}\;
\sum_{m\geq 0}\sigma^{(m)}_+\;x^{2m}\;\bar\xi^{(m)}_+\left(\frac{x}{q}\right)\cr
&+&\left(\left(1+q^2\;\epsilon_{+-}^{IR}(q)\right)\,C_-(q)
+ q^2\;\epsilon_{++}^{IR}(q)\,C_+(q)\right)\,\frac{a_-^{IR}}{q^{2\nu_{IR}}}\;
\sum_{m\geq 0}\sigma^{(m)}_-\;x^{2m+2\nu_{IR}}\;\bar\xi^{(m)}_-\left(\frac{x}{q}\right)\cr
& &
\eea
where we have introduced the holomorphic functions,
\beq\label{epsilonij}
\epsilon_{ij}^{IR}(q) \equiv \sum_{m\geq 0}\; \varphi_{ij}^{(m)}\;q^{2m}
\eeq

On the other hand, by using the series expansion of
$\xi_\infty\left(\frac{x}{q}; q\right)$ valid for $x<1$ we have,
\beq\label{xiinfexp}
z^{-\Delta_-^{IR}}\;\xi_\infty(z;q) = \sum_{m\geq 0}\;\left(\sigma^{(m)}_+\;x^{2m}+\gamma\;
\sigma^{(m)}_-\;x^{2m + 2\nu_{IR}}\right)\qquad;\qquad
\gamma\equiv \frac{\Gamma(-\nu_{IR})\,}{2^{2\,\nu_{IR}}\,\Gamma(\nu_{IR})}
\eeq
Now from (\ref{matcheq}) we have that at fixed $x< minimum (m_{IR}, 1)$,
in the limit $q\rightarrow 0\;$ equations (\ref{xiexpIR}) and (\ref{xiinfexp}) should coincide.
More specifically, if we introduce $\delta C_\pm(q)$ by,
\bea\label{deltaCpm}
C_+(q) &\equiv& \frac{1}{D(q)}\;\left(\frac{1}{a^{IR}_+}\;\left(1+q^2\;\epsilon_{+-}^{IR}(q)\right)
+\frac{\gamma}{a^{IR}_-}\;q^{2+2\nu_{IR}}\;\epsilon_{--}^{IR}(q)\right)+ \delta C_+(q)\cr
C_-(q)&\equiv& \frac{1}{D(q)}\;\left(-\frac{1}{a^{IR}_+}\;q^2\;\epsilon_{++}^{IR}(q)\;
+\frac{\gamma}{a^{IR}_-}\;q^{2\nu_{IR}}\;\left(1-q^2\;\epsilon_{-+}^{IR}(q)\right)\right)+ \delta C_-(q)\cr
&&
\eea
where,
\beq
D(q) = 1 + q^2\;\left(\epsilon_{+-}^{IR}(q)-\epsilon_{-+}^{IR}(q)\right) +
q^4\;\left(\epsilon_{++}^{IR}(q)\;\epsilon_{--}^{IR}(q)-\epsilon_{+-}^{IR}(q)\;\epsilon_{-+}^{IR}(q)\right)
\eeq
then we should get,
\bea\label{conddeltaC}
& &\lim_{q\rightarrow 0}\left\{
\sum_{m\geq 0}\sigma^{(m)}_+\,x^{2m}\;\left(\bar\xi^{(m)}_+\left(\frac{x}{q}\right)-1\right)
+\gamma\;\sum_{m\geq 0}\sigma^{(m)}_-\;x^{2m+2\nu_{IR}}\;\left(\bar\xi^{(m)}_-\left(\frac{x}{q}\right)
-1\right)\right.\cr
&+&\left.\left(\left(1-q^2\;\epsilon_{-+}^{IR}(q)\right)\,
\delta C_+(q)- q^2\,\epsilon_{--}^{IR}(q)\,\delta C_-(q)\right)\,a_+^{IR}\,
\sum_{m\geq 0}\sigma^{(m)}_+\,x^{2m}\;\bar\xi^{(m)}_+\left(\frac{x}{q}\right)\right.\cr
&+&\left.\left(\left(1+q^2\,\epsilon_{+-}^{IR}(q)\right)\delta C_-(q)
+ q^2\,\epsilon_{++}^{IR}(q)\delta C_+(q)\right)\,\frac{a_-^{IR}}{q^{2\nu_{IR}}}\,
\sum_{m\geq 0}\sigma^{(m)}_-\,x^{2m+2\nu_{IR}}\,\bar\xi^{(m)}_-\left(\frac{x}{q}\right)
\right\} =0\cr
&&
\eea
While the first line is automatically zero, the second and third lines should be zero
separately because they present different power series
\footnote{
A subtlety (not present in the case considered in the text) arises if $\bar\xi^{(m)}_-(z)$
contains powers of the form $z^{-2\nu_{IR} - 2n}$ with $n\in\aleph$; in that case
it can be easily showed that the effect is that the coefficients of $\delta C^\pm(q)$ on
the second line of (\ref{conddeltaC}) get modified by holomorphic functions;
this fact does not modify the subsequent arguments.
}.
From the third line we get,
\beq
\delta C_-(q)\xrightarrow{q\rightarrow 0} -\varphi^{(0)}_{++}\;q^2\,\delta C_+(q) + q^{2\nu_{IR}}\;A(q)
\eeq
where $A(q)\xrightarrow{q\rightarrow 0} 0$.
Then the second line of (\ref{conddeltaC}) yields,
\beq\label{deltabeha}
\delta C_+(q)\xrightarrow{q\rightarrow 0} \varphi^{(0)}_{--}\;q^{2\nu_{IR} +2}\;A(q) \quad\Rightarrow\quad
\delta C_-(q)\xrightarrow{q\rightarrow 0} q^{2\nu_{IR}}\;A(q)
\eeq
Going to (\ref{deltaCpm}) with (\ref{deltabeha}) we get the leading behaviors,
\beq\label{Cpm0sn}
C_+(0) = \frac{1}{a_+^{IR}}\qquad;\qquad
C_-(q)|_{q\rightarrow 0} = -\frac{\bar\varphi_{++}^{(0)}}{a_+^{IR}}\;q^2
\eeq
This yields for the two-point function (\ref{G21}) the Goldstone pole,
\beq\label{G22}
G_2(q)\xrightarrow{q\rightarrow 0} \frac{\alpha}{q^2}
\eeq
where by using (\ref{epsilonppm}) and (\ref{Fijm}), i.e.
\beq\label{varphi++}
\bar\varphi_{++}^{(0)} =\int_{z_i}^\infty dw\;w^{1-d}\; \xi_+(w)^2
\eeq
we get for the residue,
\beq
\label{alpha}
\alpha = \frac{2\,\nu_{UV}\,(a_+^{UV})^2}{\int_{0}^\infty dw\;w^{1-d}\; \xi_+(w)^2}
\eeq
The result is reassuring in the sense that both contributions in the denominator
of (\ref{G21}) add to yield a $z_i$-independent result.

\section{\label{global}Global symmetries and AdS/CFT}

Let us now use all this machinery for a simple d-dimensional strongly coupled system with a spontaneously
broken global symmetry. We would like to see explicitly what makes Nambu-Goldstone bosons massless
in the AdS/CFT picture: it is the square integrability of the solution $\xi_+(z)$ of the perturbation equation.
In other words, a normalizable perturbation is massless.

The simplest example seems to be SU(3) $\to$ SU(2) $\times$ U(1). A physically more appealing case could
be SU(5) $\to$ SU(3) $\times$ SU(2) $\times $U(1). The hope is that eventually one could then weakly couple the
system to gauge bosons, i.e. gauge it. Let's consider a real adjoint, which we parametrize as

\bea
\Sigma=\frac{1}{\sqrt{2}}
\bem
t/\sqrt{3}+t_3 & t_1-i t_2 & w_1-i w_2 \cr
t_1+i t_2 & t/\sqrt{3}-t_3 & w_3-i w_4 \cr
w_1+i w_2 & w_3+i w_4 & -2t/\sqrt{3}
\eem
\eea

\noi
and a complex fundamental:

\beq
\bar F=
\bem
T_1 & T_2 & H
\eem^T
\eeq

Let the superpotential be

\beq
W=\Delta\left(\frac{1}{2}Tr\Sigma^2+\frac{\sqrt{6}}{3}Tr\Sigma^3\right)
+F^\dagger\left(m-\sqrt{6}\alpha\Sigma\right)F
\eeq

The strange relation between the $\Sigma^2$ and $\Sigma^3$
coefficients are chosen so that the straightforward generalization of the potential (\ref{Vatkappa0})
\bea
V&=&\frac{1}{2}\left(\frac{\partial W}{\partial t}\right)^2+
\frac{1}{2}\sum_{i=1}^{3}\left(\frac{\partial W}{\partial t_i}\right)^2+
\frac{1}{2}\sum_{a=1}^{4}\left(\frac{\partial W}{\partial w_a}\right)^2\nonumber\\
&+&\sum_{i=1}^{2}\left(\frac{\partial W}{\partial T_i}\right)\left(\frac{\partial W}{\partial T_i^*}\right)+
\left(\frac{\partial W}{\partial H}\right)\left(\frac{\partial W}{\partial H^*}\right)-d\;W
\eea

\noi
has an extremum at $t=0$ and a minimum at $t=1$ with all other fields vanishing, and the
potential for $t$ is the same as in \cite{Bajc:2013wha}:

\bea
W(t)&=&\Delta\left(\frac{t^2}{2}-\frac{t^3}{3}\right)\\
V(t)&=&\frac{1}{2}W'^2(t)-dW(t)\nonumber\\
&=&\Delta(\Delta-d)\frac{t^2}{2}-\Delta(3\Delta-d)\frac{t^3}{3}+2\Delta^2\frac{t^4}{4}
\eea

The solution to the e.o.m. is

\beq
t(z)=\frac{z^\Delta}{1+z^\Delta}\;\;,\;\;t_i,w_a=0
\eeq

One can calculate the mass matrix

\bea
\frac{\partial V}{\partial t^2}&\equiv&m^2(t)=
\Delta(\Delta-d)-2\Delta(3\Delta-d)t+6\Delta^2t^2\\
\frac{\partial V}{\partial w_a\partial w_b}&\equiv&m^2_w(t)\delta^{ab}=
\left(\Delta(\Delta-d)-\Delta(3\Delta-d)t+2\Delta^2t^2\right)\delta^{ab}\\
\frac{\partial V}{\partial t_i\partial t_j}&\equiv&m^2_t(t)\delta^{ij}=
\left(\Delta(\Delta-d)+2\Delta(3\Delta-d)t+2\Delta^2t^2\right)\delta^{ij}\\
\frac{\partial V}{\partial T^*_\alpha\partial T_\beta}&\equiv& m_T^2(t)\delta^{\alpha\beta}=
\left(m(m-d)-\alpha(2m+\Delta-d)t+\alpha(\alpha+\Delta)t^2\right)\delta^{\alpha\beta}\\
\frac{\partial V}{\partial H^*\partial H}&\equiv& m_H^2(t)=
m(m-d)+2\alpha(m+\Delta-d)t+2\alpha(2\alpha-\Delta)t^2
\eea

\noi
with all other elements vanishing.

This means that it is easy to solve the perturbation equation since the different modes decouple. In an
obvious notation:

\bea
\label{xi}
z^2\;\ddot{\xi}(z;q)-(d-1)\;z\;\dot{\xi}(z;q)-\left(q^2\,z^2 + m^2(t(z))\right)\;\xi(z;q)&=&0\\
\label{xiw}
z^2\;\ddot{\xi}^w(z;q)-(d-1)\;z\;\dot{\xi}^w(z;q)-\left(q^2\,z^2 + m_w^2(t(z))\right)\;\xi^w(z;q)&=&0\\
\label{xit}
z^2\;\ddot{\xi}^t(z;q)-(d-1)\;z\;\dot{\xi}^t(z;q)-\left(q^2\,z^2 + m_t^2(t(z))\right)\;\xi^t(z;q)&=&0\\
\label{xiT}
z^2\;\ddot{\xi}^T(z;q)-(d-1)\;z\;\dot{\xi}^T(z;q)-\left(q^2\,z^2 + m_T^2(t(z))\right)\;\xi^T(z;q)&=&0\\
\label{xiH}
z^2\;\ddot{\xi}^H(z;q)-(d-1)\;z\;\dot{\xi}^H(z;q)-\left(q^2\,z^2 + m_H^2(t(z))\right)\;\xi^H(z;q)&=&0
\eea

The first equation (\ref{xi}) has a well known solution

\beq
\xi_+(z)=z\;\dot{t}(z)
\eeq

The second one (\ref{xiw}) is for the Goldstone-bosons of the global symmetry. Since

\beq
m_w^2(t)=\frac{1}{t}V'(t)
\eeq

\noi
the well-behaved solution is simply

\beq
\xi^w_+(z)=t(z)
\eeq

This is why it has a pole at $q^2=0$. We just need to do the usual expansion derived in general in
the previous section, see also \cite{Bajc:2013wha}, with the result for the propagator

\beq
G_2^w(q)=\frac{\alpha}{q^2}
\eeq

\noi
with the general expression

\beq
\alpha=\frac{2\nu_{UV}\left(a_+^{UV}\right)^2}{(\int_0^\infty dx x^{1-d}\left(\xi_+^w(x)\right)^2}
\eeq

In our specific case (\ref{xiw}) we have

\beq
a_+^{UV}=1
\eeq

\noi
with the integral in the denominator finite.

Then, what about the third equation (\ref{xit}), i.e. for $\xi^t$? One can easily find the solution for $q=0$:

\beq
\xi^t(z)=C_1z^\Delta(1+z^\Delta)^2+C_2\frac{z^{d-\Delta}}{1+z^\Delta}
\;_2F_1(1,-5+d/\Delta,-1+d/\Delta,-z^\Delta)
\eeq

For $z\to\infty$ we get

\beq
\xi^t(z)\to C_1z^\Delta(1+z^\Delta)^2+C_2\frac{z^{d-\Delta}}{1+z^\Delta} z^{5\Delta-d}
\frac{\Gamma(6-d/\Delta)\Gamma(-1+d/\Delta)(1+z^\Delta)^3}{2z^{3\Delta}}
\eeq	

\noi
and so

\beq
\label{C1}
C_1=-\frac{C_2}{2}\Gamma(6-d/\Delta)\Gamma(-1+d/\Delta)
\eeq

In the opposite limit $z\to0$ (\ref{xit}) becomes

\beq
\xi^t(z)\to C_1\left(z^\Delta+\ldots\right) +C_2\left(z^{d-\Delta}+\ldots\right)
\eeq

\noi
so that due to (\ref{C1}) we get in the IR limit $q\to 0$

\beq
G_2^t(0)=\frac{C_1}{C_2}=-\frac{1}{2}\Gamma(6-d/\Delta)\Gamma(1-d/\Delta)
\eeq

Obviously there is no pole here at $q=0$, a pole is expected at finite $q$. The reason
for no pole at $q=0$ is thus due to the fact that there is no solution finite in the whole
positive $z$-axis. This was true for $\xi(z)=z\;\dot{t}(z)$ and $\xi^w(z)=t(z)$, and this is why the
next order in $q$ was needed there. In other words, if the integral in (\ref{alpha}) is finite, the
propagator obeys (\ref{G22}), if it is not, then the leading term in this expansion is a constant.

Equations (\ref{xiT}) and (\ref{xiH}) for $\xi^T$ and $\xi^H$ seem to point to the same conclusion as
for $\xi^t$: no pole, i.e. no light degree of freedom. So if we would like one of the two to be light, i.e.
for example $\xi^H$ (the analog of the light SM doublet in SU(5)), we would need to further fine-tune
the system, similarly as one obtains the usual doublet-triplet splitting in a SU(5) grand unified theory.

\section*{Acknowledgments}
We would like to thank Mirjam Cveti\v c, Fidel Schaposnik and Guillermo Silva for discussions,
and Mirjam Cveti\v c, Carlos Hoyos, Uri Kol, Cobi Sonnenschein and Shimon Yankielowicz
for correspondence. This work has been supported in part by the Slovenian Research
Agency, and by the Argentinian-Slovenian programme BI-AR/12-14-004 //
MINCYT-MHEST SLO/11/04.

\end{document}